\documentclass[9pt, conference]{IEEEtran}
\IEEEoverridecommandlockouts

\usepackage{amsmath,amssymb,amsfonts}
\usepackage{algorithmic}
\usepackage{graphicx}
\usepackage{textcomp}
\usepackage{xcolor}
\usepackage{url}
\usepackage[style=ieee]{biblatex}

\begin{document}
\title{Towards An Integrated Approach for Expressive Piano Performance Synthesis from Music Scores\\

\thanks{This work is supported by both the UKRI Centre for Doctoral Training in Artificial Intelligence
and Music [grant number EP/S022694/1], and the National Institute of Informatics in Japan. J.Tang is a research student supported jointly by the China Scholarship Council [grant number 202008440382] and Queen Mary University of London.  E. Cooper conducted this work while at NII, Japan and is currently employed by NICT, Japan.}
}

\author{
\IEEEauthorblockN{Jingjing Tang\IEEEauthorrefmark{1}, 
Erica Cooper\IEEEauthorrefmark{2}, 
Xin Wang\IEEEauthorrefmark{2}, 
Junichi Yamagishi\IEEEauthorrefmark{2}, 
Gy\"orgy Fazekas\IEEEauthorrefmark{1}}
\IEEEauthorblockA{\IEEEauthorrefmark{1}Centre for Digital Music, Queen Mary  University of London, UK}
\IEEEauthorblockA{\IEEEauthorrefmark{2}National Institute of Informatics, Japan}
\IEEEauthorblockA{Email: jingjing.tang@qmul.ac.uk, george.fazekas@qmul.ac.uk, ecooper@nict.go.jp, wangxin@nii.ac.jp, jyamagis@nii.ac.jp}}
\maketitle
\begin{abstract}
This paper presents an integrated system that transforms symbolic music scores into expressive piano performance audio. By combining a Transformer-based Expressive Performance Rendering (EPR) model with a fine-tuned neural MIDI synthesiser, our approach directly generates expressive audio performances from score inputs. To the best of our knowledge, this is the first system to offer a streamlined method for converting score MIDI files lacking expression control into rich, expressive piano performances. We conducted experiments using subsets of the ATEPP dataset, evaluating the system with both objective metrics and subjective listening tests. Our system not only accurately reconstructs human-like expressiveness, but also captures the acoustic ambience of environments such as concert halls and recording studios. Additionally, the proposed system demonstrates its ability to achieve musical expressiveness while ensuring good audio quality in its outputs.
\end{abstract}

\begin{IEEEkeywords}
expressive performance generation, music audio synthesis, music generation, deep learning, Transformer
\end{IEEEkeywords}

\section{Introduction} \label{sec:intro}
Expressive Performance Rendering (\textbf{EPR}) aims to replicate human-like musical interpretations using computational systems that take symbolic music scores as input. Research in this field primarily focuses on systems interpreting music scores in formats such as MIDI or MusicXML \cite{cancino2018computational}. These systems aim to generate expressive performance MIDI files by enriching mechanical score MIDI files, which lack the nuanced dynamics, articulation, and variations necessary to convey emotion and expression. Using symbolic representations has enabled the adaptation of NLP models, particularly Transformers, in EPR systems \cite{borovik2023scoreperformer, worrall2024comparative}. Additionally, various tokenisation methods \cite{borovik2023scoreperformer, hsiao2021compound, chou2021midibert, tang2023reconstructing} have been developed to apply Transformers to symbolic music generation tasks, including EPR.

Numerous studies employing deep learning techniques have demonstrated convincing outcomes. These include methods such as Recurrent Neural Networks (RNN) \cite{Jeong2019VirtuosoNetAH}, Graph Neural Networks (GNN) \cite{jeong2019graph}, Generative Adversarial Networks (GAN) \cite{renault2023expressive}, and diffusion models \cite{zhang2024dexter}. Transformers \cite{borovik2023scoreperformer, rhyu2022sketching} have further been utilised for generating controlled expressions, particularly in modulating dynamics and tempo in line with specific performance directives embedded into the model. While large-scale transcribed MIDI performance datasets \cite{zhang2022atepp, edwards2023pijama, kong2020giantmidi} are available, their widespread use is limited due to the need for highly accurate annotations. The lack of the precise score-performance alignments necessary for detailed analysis hinders their full utilisation in relevant studies. Another recent work \cite{worrall2024comparative} has employed multiple Transformer models to predict different note-wise features, relying solely on the note's onset and offset times extracted from music scores. In contrast, our approach adopts a more streamlined system architecture inspired by our previous work aiming at reconstructing human expressiveness in transcribed scores \cite{tang2023reconstructing}. We utilise a single Transformer encoder, which boosts computational efficiency and requires only note onset and offset time for generating the performances. This design enables the use of transcribed piano performance datasets to directly render expressive piano performances from mechanical scores\footnote{Demo and codes: \url{https://tangjjbetsy.github.io/S2A/}}.
\begin{figure}
\centering
  \includegraphics[width=.75\columnwidth]{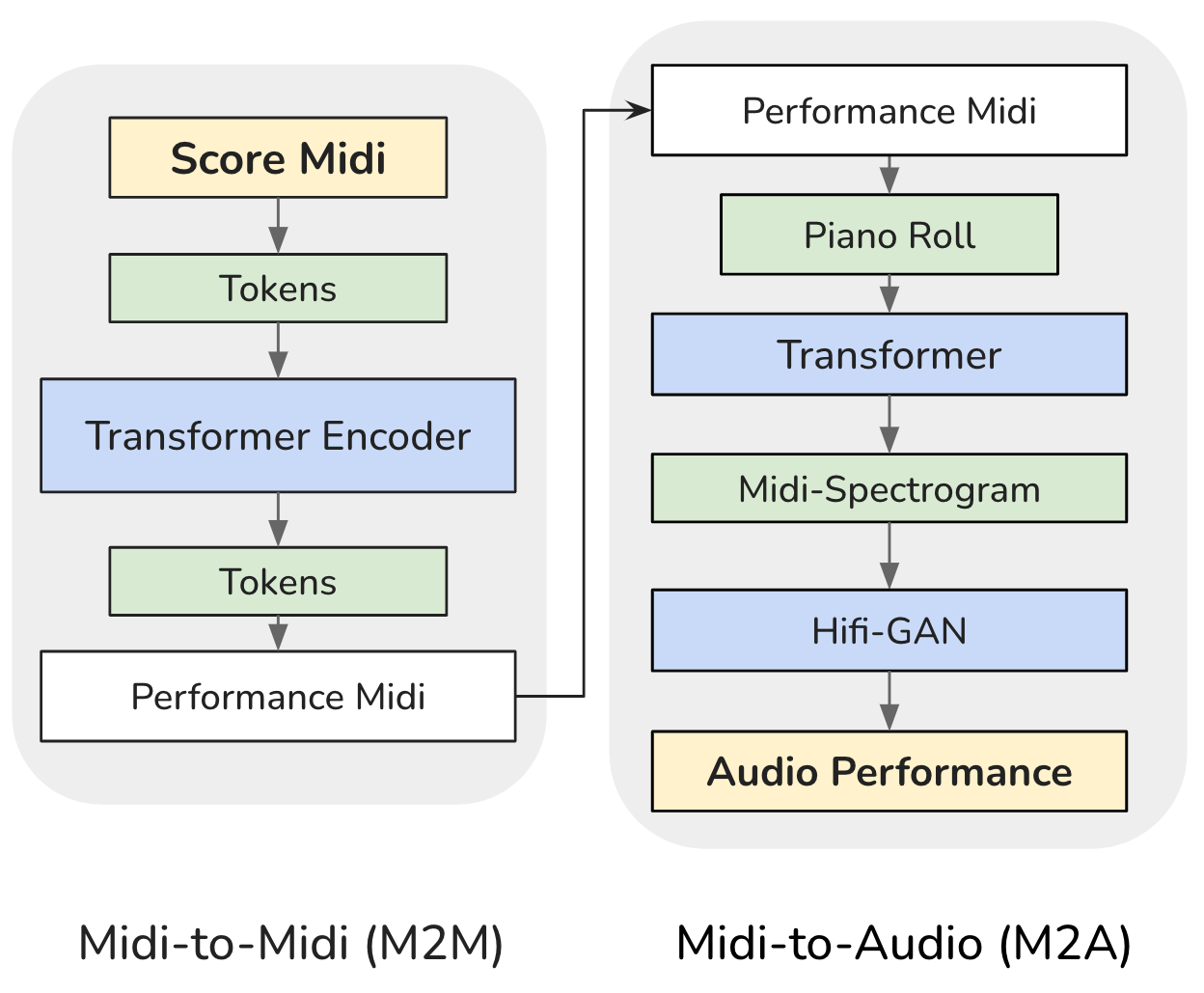}
  \caption{Overview of the proposed score-to-audio system. The left part (\textbf{M2M}) illustrates our EPR model, featuring a Transformer encoder. The right section (\textbf{M2A}) details the architecture of the MIDI synthesiser, which incorporates a Transformer model \cite{li2019neural} adapted from text-to-speech (TTS) tasks and a HiFi-GAN vocoder \cite{kong2020hifi}.}
  \label{fig:overview}
  \vspace{-1em}
\end{figure}

While the majority of existing research in this domain has focused on MIDI-to-MIDI generation, the innovative MIDI-DDSP \cite{wu2021midi} system has introduced a method for producing audio outputs from mechanical score MIDI inputs. Unlike MIDI-DDSP, which is trained on multi-instrument datasets, but limits each instrument's performance to monophonic outputs, our work focuses on the classical piano, presenting a more intricate challenge due to its inherently polyphonic nature. Inspired by the MIDI-DDSP system, we have developed an integrated solution that combines the proposed Transformer-based MIDI-to-MIDI (\textbf{M2M}) EPR model with a fine-tuned neural MIDI synthesiser \cite{shi2023can}, which is a MIDI-to-Audio (\textbf{M2A}) model. The integrated system first transforms mechanical score MIDI into expressive performance MIDI using our M2M model, and then directly synthesises this performance MIDI into high-quality audio performances using our MIDI synthesiser. Differing from sample-based synthesis methods such as Fluidsynth \cite{newmarch2017fluidsynth} and physical modelling techniques like Pianoteq\footnote{\url{https://www.modartt.com/pianoteq}}, the data-driven MIDI synthesiser facilitates a more straightforward reconstruction of timbre across various pianos and recording environments. This approach adapts the model design from the domain of text-to-speech synthesis \cite{shi2023can}. The proposed approach, assisted by the synthesiser, simplifies the integration of expressive control and audio synthesis, enabling efficient generation of expressive performance audio. To our knowledge, this is the first system to seamlessly convert mechanical score MIDI files into expressive piano performance audio using purely deep learning models.

\section{Proposed Method}
\begin{figure}
\centering\includegraphics[width=.75\columnwidth]{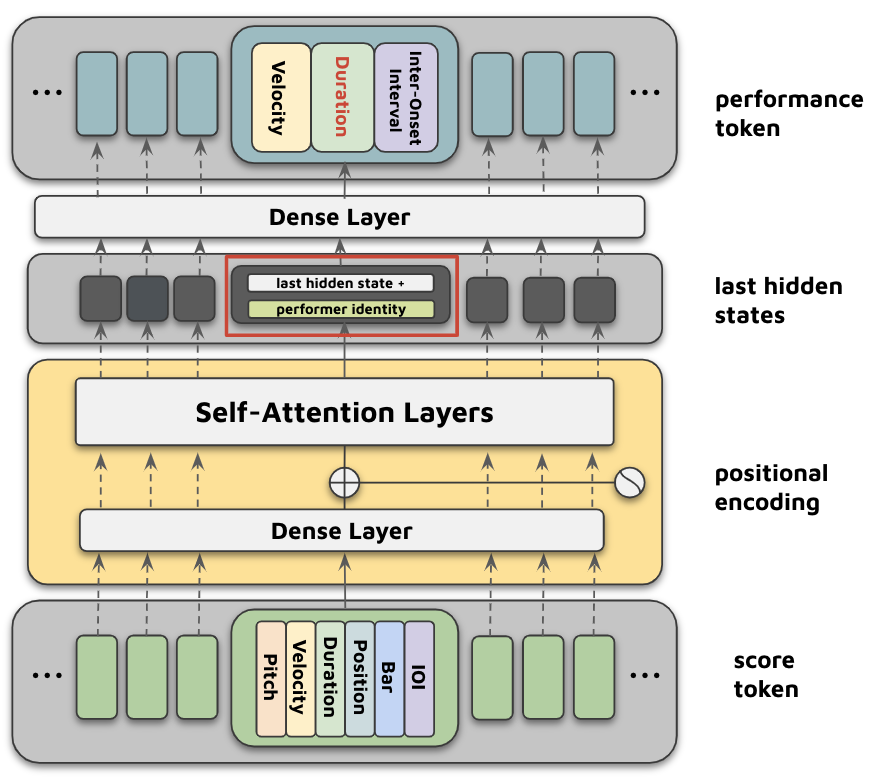}
  \caption{Architecture of the M2M model, adapted from our previous work~\cite{tang2023reconstructing} Modifications to the original design are highlighted in red font and enclosed in red boxes for clarity.}
  \label{fig:architecture}
  \vspace{-1em}
\end{figure}
\begin{table}
    \caption{Vocabulary size of the tokenised note-level features}
    \label{tab:token_numbers}
    \small
    \centering
    \begin{tabular}{l|cccccc}
    \hline 
    \textbf{Feature} & \textbf{Pitch} & \textbf{Vel.} & \textbf{Dur.} & \textbf{IOI} & \textbf{Pos.} & \textbf{Bar} \\ \hline
    \textbf{Size} & 92 & 68 & 1156 & 772 & 388 & 3004 \\
    \hline
    \end{tabular}
    \vspace{-1em}
\end{table}
The overview of our system is shown in Fig.~\ref{fig:overview}, which illustrates the pipeline for rendering audio performances from score inputs. Our system integrates two models: a Transformer-based M2M model for generating expressive performance MIDI files, and a MIDI synthesiser for producing the corresponding audio outputs. While the M2M model is trained from scratch, the MIDI synthesiser, initially trained on the Maestro dataset \cite{shi2023can}, is fine-tuned for compatibility with the M2M model. Additionally, we create a baseline based on the M2A model by substituting the performance MIDI with the score MIDI as input.
\subsection{M2A and M2M Alignment}\label{sec:aligment}
We used an alignment algorithm by Nakamura et al.~\cite{Nakamura2017PerformanceED} to produce note-wise alignments between the performance MIDI and the score MIDI. The transcribed performance MIDI aligns closely with the recordings in the ATEPP \cite{zhang2022atepp} dataset, resulting in aligned triplets of score MIDI, performance MIDI, and audio recordings. This alignment facilitates the fine-tuning of the M2A model and the creation of the baseline model.
\subsection{Fine-tuning the M2A Model}
\label{sec:fine-tune}
Unlike the Maestro dataset \cite{hawthorne2018enabling}, which consists of competition recordings captured with Disklavier pianos, the ATEPP \cite{zhang2022atepp} dataset encompass a diverse range of recordings from various professional settings and live concerts, all transcribed using deep learning algorithms. To address the knowledge gap, we jointly fine-tuned the Transformer model \cite{li2019neural} and the HiFi-GAN vocoder \cite{kong2020hifi} using audio and performance MIDI pairs. As outlined in the paper \cite{shi2023can}, the Transformer model was designed to transform the MIDI piano rolls into MIDI-scale spectrograms \cite{cooper2021text}, while the vocoder created the waveforms based on the spectrograms. All of the performances were cut into segments of 9.6 seconds, with concatenation applied as a post-processing step during the generation phase.
\subsection{Baseline Construction}
\label{sec:baseline}
We proposed a baseline model designed to exhibit musical expressiveness and generate audio outputs through a single-stage process, serving as a comparison to the two-stage method. This baseline shares a similar structure to the M2A model, but incorporates performer and album identity by adding identity embeddings to the Transformer's final hidden state. Unlike the M2A model, the baseline uses score MIDI as input instead of performance MIDI. To leverage the acoustic ambience features already learned, the training parameters of the baseline model were initialized using the fine-tuned M2A model.
\subsection{Midi Tokenisation}
\label{sec:tokenisation}

We adapted the Octuple tokenisation method~\cite{zeng-etal-2021-musicbert} to create aligned performance and score MIDI tokens for six distinct note-wise features: pitch, velocity (\textbf{Vel.}), duration (\textbf{Dur.}), inter-onset interval (\textbf{IOI}), bar, and position (\textbf{Pos.}). In the score MIDI files, velocity was set to a constant value of 60 to eliminate expressiveness and enable the use of MIDI converted from score XML which typically lacks velocity data. This approach improves training by reducing the overall vocabulary size, which refers to the number of unique tokens the model needs to learn, while simultaneously capturing distinct features with tailored vocabularies. Following our previous work \cite{tang2023reconstructing}, we define IOI similarly but adopt a beat resolution of 96, cutting the vocabulary size to a quarter of the earlier study. Unlike the previous work, we tokenise the IOI values, with the resulting vocabulary detailed in Table~\ref{tab:token_numbers}. The performance and score MIDI sequences were segmented into 256-note sequences for training the M2M model.
\subsection{M2M Model}
Modifications were made to the architecture developed in our previous work~\cite{tang2023reconstructing}, as depicted in Fig.~\ref{fig:architecture}. Instead of linearly concatenating performer identity embedding to the last hidden state, we created embedding of matching dimensions and summed it with the last hidden state, enhancing the representation of pianist identity. This approach is inspired by techniques used for speaker embedding in text-to-speech synthesis \cite{li2019neural, chen2020multispeech}. Instead of predicting deviations in note duration, we opted to predict the actual duration values, resulting in a more straightforward and direct system. This result, combined with the predicted velocity and IOI values, enables us to reconstruct a performance MIDI using the given pitches. Additionally, we employed a probabilistic loss function, replacing the previously used deterministic one, and applied temperature and nucleus sampling \cite{hsiao2021compound} during generation to enhance output variety.
\begin{table*}
    \centering
    \caption{Performance metrics for the M2M model: means and 95\% confidence intervals}
    \small
    \begin{tabular}{l|ccc|ccc}
        \hline
        & \multicolumn{3}{c|}{\textbf{\textit{ Performance-wise}}} & \multicolumn{3}{c}{\textit{\textbf{Segment-wise}}} \\
        \cline{2-7}
        \textbf{Feature} & \textbf{KLD $\downarrow$} & \textbf{Correlation $\uparrow$} & \textbf{DTWD $\downarrow$} & \textbf{KLD $\downarrow$} & \textbf{Correlation $\uparrow$} & \textbf{DTWD $\downarrow$} \\
        \hline
        Velocity & 0.018 $\pm$ 0.001 & 0.831 $\pm$ 0.017 & 0.063 $\pm$ 0.002 & 0.016 $\pm$ 0.001 &  0.665 $\pm$ 0.011 & 0.064 $\pm$ 0.001\\
        Inter-Onset Interval & $<$ 0.001 & 0.990 $\pm$ 0.003 & 0.010 $\pm$ 0.001 & $<$ 0.001 & 0.932 $\pm$ 0.006 & 0.009 $\pm$ 0.001 \\
        Duration & 0.187 $\pm$ 0.010 & 0.753 $\pm$ 0.017 & 0.026 $\pm$ 0.003 & 0.184 $\pm$ 0.005 & 0.668 $\pm$ 0.012 & 0.023 $\pm$ 0.001 \\
    \hline
    \end{tabular}
    \label{tab:M2M_matrics}
    \vspace{-.8em}
\end{table*}
\section{Experiments}
\subsection{Experiment  Data}\label{sec:datasets}
All our experiments utilise subsets of ATEPP-1.2 \footnote{\url{https://github.com/tangjjbetsy/ATEPP}}, the latest version of a large-scale transcribed piano performance dataset \cite{zhang2022atepp}. We selected two distinct subsets to achieve different objectives: subset \textbf{A} was used to fine-tune the MIDI synthesiser, bridging the knowledge gap between the Maestro and ATEPP datasets, while subset \textbf{B} was used to train the M2M model and construct the baseline. 

Subset \textbf{A} includes 371 performances from the top 9 pianists and 5 composers, sourced from 75 different albums across diverse acoustic environments. Subset \textbf{B} focuses primarily on Beethoven sonatas, along with one composition by Mozart, allowing for stylistic comparisons among different performers. The selection was constrained by the availability of scores within the dataset, resulting in 735 performances across 40 different compositions. There is no performances overlap between the two subsets. Subset \textbf{A} was split into training and validation sets with a 9:1 ratio, while subset \textbf{B} was divided into training, validation, and testing sets with an 8:1:1 ratio based on the number of performances per performer.

\subsection{Implementation}
The M2A model, with approximately 102M parameters, was fine-tuned over 200 epochs using subset \textbf{A} on four Tesla A100 GPUs, following the methodologies in \cite{shi2023can} and its open-source implementation\footnote{\scriptsize\url{https://github.com/nii-yamagishilab/midi-to-audio}}. Based on this fine-tuned model, we constructed the baseline by training for an additional 200 epochs with subset \textbf{B}.

The M2M model, with around 12M parameters, was trained on subset \textbf{B}. Training spanned up to 2000 epochs with a dynamic batch size of 256, using the Adam optimizer with an initial learning rate of 2e-5 and a 40-step warm-up scheduler to improve convergence. Training was completed within 4 hours on a Tesla A100 GPU. We employed the GradNorm algorithm \cite{pmlr-v80-chen18a} to dynamically adjust feature loss weights, ensuring balanced learning across features.
\section{Evaluation}
We selected different objective metrics and conducted independent listening tests to assess the performance of the M2M and M2A models, recognising their unique architectural designs and specific objectives. All metrics in objective evaluations were calculated using the test set from subset \textbf{B} to ensures consistency.
\subsection{Objective Evaluation for M2M Model}
The quality of the intermediate outputs produced by the M2M model was evaluated using three metrics: Kullback-Leibler divergence (\textbf{KLD}), Pearson correlation, and dynamic time warping distance (\textbf{DTWD}), as summarized in Table~\ref{tab:M2M_matrics}. The first two metrics, proposed in prior work \cite{borovik2023scoreperformer, zhang2024dexter}, evaluate EPR system outputs. KLD measures the divergence between predicted and target performance distributions for each feature, while Pearson correlation assesses the relationship between predicted and target feature sequences. Additionally, we consider DTWD as another approach to measure the similarity between the predictions and the ground truth. Given that polyphonic music involves simultaneous notes, resulting in varying note orders for chords during data processing, DTWD is assumed to effectively capture chord similarity in the performances regardless of note order. In our experiments, DTWD values are averaged across notes and normalised by the corresponding vocabulary sizes for clearer comparison, with lower values indicating a closer match. 

Due to architectural constraints, the current Transformer encoder in the M2M model processes a maximum of 256 notes at a time. Segments are then concatenated to form complete performances. To evaluate the effect of such post-processing, we calculated the metrics for full performances and 256-note segments. As presented in Table~\ref{tab:M2M_matrics}, the three metrics are calculated on the test set. The means are provided with 95\% confidence intervals (CIs) across full performances and segments of 256 notes. Our results indicate consistently strong performance for both complete performances and segments, demonstrating that post-concatenation does not adversely affect model outcomes. However, the lower correlation in segment evaluations indicates that full performance assessments may overlook occasional anomalies in specific local sections. The reconstruction of IOI and velocity is quite effective, while the prediction of duration still has room for improvement.
\subsection{Objective Evaluation for M2A Model}
\label{sec:oe_m2a}
\begin{table}
    \centering
    \caption{Performance metrics for the M2A Model and Baseline}
    \small
    \begin{tabular}{c|cc}
        \hline
        \textbf{Systems} & \textbf{Chroma $\downarrow$} & \textbf{Spectrogram $\downarrow$} \\
        \hline        Pianoteq&\textbf{0.487$\pm$0.008}&0.294$\pm$0.013\\ 
        \hline        Baseline&0.624$\pm$0.027&0.284$\pm$0.013\\
        M2A \cite{shi2023can} &0.539$\pm$0.021&0.318$\pm$0.013\\
        Fine-tuned M2A (\textbf{ours})&0.522$\pm$0.018&\textbf{0.262$\pm$0.009}\\
    \hline
    \end{tabular}
    \label{tab:M2A_matrics}
    \vspace{-1em}
\end{table}

To assess the fine-tuned M2A model, we computed Chroma and MIDI spectrogram mean square errors (MSEs) as described in the original paper \cite{shi2023can}. The same evaluation was applied to assess the M2A model both with and without fine-tuning, and in comparison to the physical model Pianoteq \cite{pianoteq}. Additionally, we calculated the metrics for the baseline model, considering its similar structural design. However, it is important to note that these metrics may not fully capture the extent of expressiveness added to the input scores.

As shown in Table~\ref{tab:M2A_matrics}, the means and 95\% confidence intervals are calculated across performances in the test set. The results show that fine-tuning enhanced the M2A model's performance on the ATEPP dataset. When considering ambient sounds, as measured by spectrogram distance, the outputs of the fine-tuned model closely aligns with the targets. However, inconsistencies were observed in acoustic ambience for complete performances due to the architecture's limitation of generating audio in segments of up to 9.6 seconds each, as discussed in Sec.~\ref{sec:fine-tune}. The Chromagram results, which assess the accuracy of musical key reconstruction, indicate that the fine-tuned model is less accurate compared to Pianoteq. Segmentation artefacts were minimised in the evaluation by selecting the optimal concatenation points using cross-correlation to ensure smoother transitions between segments.

Compared to the M2A model without fine-tuning, the baseline model exhibited a larger Chroma distance and a smaller spectrogram distance. The results indicate that although it may be possible to take into account the recording environment and the distinct sounds of different piano manufacturers, there are still issues with accurately reproducing the sound of each key. Further analysis of the outputs revealed that the primary issue stem from the corrupted MIDI-scale spectrograms produced by the Transformer. We believe this limitation is due to the Transformer's constrained ability to simultaneously learn musical expressiveness and acoustic features, given the relatively small training set of approximately 65 hours of piano performances in subset \textbf{B}.

\begin{table}
  \centering
  \caption{Mean opinion scores (MOS) for expressiveness and quality in the evaluated systems from the two listening tests}
  \small
  \begin{tabular}{l|c}
  \hline
  \hline
  \multicolumn{2}{l}{\textbf{Test A:} \textit{Evaluating the M2M Model: \textbf{Expressiveness}}} \\
  \hline
  \textbf{Systems (Midi Source + Synthesiser)} & \textbf{MOS}\\
  \hline
  S1. Groundtruth (GT.) + Pianoteq (Ref.) & 8.58 $\pm$ 0.41\\
  \hline
  S2. M2M Output (\textbf{ours}) + Pianoteq & 6.69 $\pm$ 0.49\\
  S3. M2M Output + Fine-tuned M2A (\textbf{ours}) & 3.87 $\pm$ 0.57\\
  S4. Score + Pianoteq & 5.07 $\pm$ 0.58\\ 
  S5. Score + Baseline & 1.54 $\pm$ 0.38\\
  \hline
  \hline
  \multicolumn{2}{l}{\textbf{Test B:} \textit{Evaluating the M2A Model: \textbf{Quality}}} \\
  \hline
  \textbf{Systems (Midi Source + Synthesiser)} & \textbf{MOS}\\
  \hline
  S0. Human Performance Recording (Ref.)&7.19 $\pm$ 0.39\\
  \hline
  S1. Groundtruth + Pianoteq&7.41 $\pm$ 0.40\\
  S6. Groundtruth + M2A~\cite{shi2023can}&6.29 $\pm$ 0.49\\
  S7. Groundtruth + Fine-tuned M2A (\textbf{ours})&5.96 $\pm$ 0.47\\
  S3. M2M Output + Fine-tuned M2A (\textbf{ours}) &5.27 $\pm$ 0.52\\
  S5. Score + Baseline&5.12 $\pm$ 0.59\\
  \hline
  \end{tabular}
  \label{tab:listening_test}
  \vspace{-1em}
\end{table}

\subsection{Subjective Evaluation}
Two separate listening tests were conducted to evaluate the performance of the M2M (\textbf{Test A}) and M2A (\textbf{Test B}) models respectively. In total, 34 participants were recruited for the listening tests: 13 participants evaluated the perceived expressiveness of the M2M model's output, while 21 participants assessed the audio quality of the fine-tuned M2A model's outputs. Nearly all participants had some level of music training, either in piano or other musical instruments.

Four compositions by Beethoven were selected, with two compositions used during model training (\textbf{internal}) and two which were unseen during training (\textbf{external}). From each composition, four segments of 10-15 seconds were randomly extracted. For the M2M model evaluation, a total of 12 segments (3 per composition) were used, while 16 segments (4 per composition) were utilised for the M2A model evaluation, reflecting the number of participants. The systems (\textbf{S0-S7}) evaluated in each test, along with their mean opinion scores (\textbf{MOS}) and corresponding 95\% CIs, are detailed in Table~\ref{tab:listening_test}. In both tests, each stimulus was rated by at least five participants.

In the first test (\textbf{Test A}), participants rated the \textit{expressiveness} of the stimuli on a scale from 0 to 100 (scaled to 10 here), focusing on how expressive, natural, and human-like the performances were in comparison to the reference human performance (\textbf{S1}). We particularly asked the participants to ignore the audio quality and acoustic environments. 
The results indicate that although the outputs of the M2M model (\textbf{S2}) were rated lower than the reference human performance (\textbf{S1}), they were still perceived as more expressive than the mechanical scores (\textbf{S4}), demonstrating the model's success in reconstructing expression. However, the fine-tuned M2A model (\textbf{S3}) exhibited a reduced level of expressiveness compared to Pianoteq, possibly due to the absence of pedalling in the M2M outputs. Although the proposed integrated system (\textbf{S3}) was rated lower in expressiveness compared to the synthesised mechanical scores (\textbf{S4}), it is important to note that participants' perceptions may have been influenced by the audio quality, despite being instructed to disregard it. The rating for the baseline model (\textbf{S5}) highlights its lack of success in adding expressiveness to the scores, which may also be affected by the key sound reproduction issues discussed in Section~\ref{sec:oe_m2a}.

\begin{figure}
\centering
\includegraphics[width=\columnwidth]{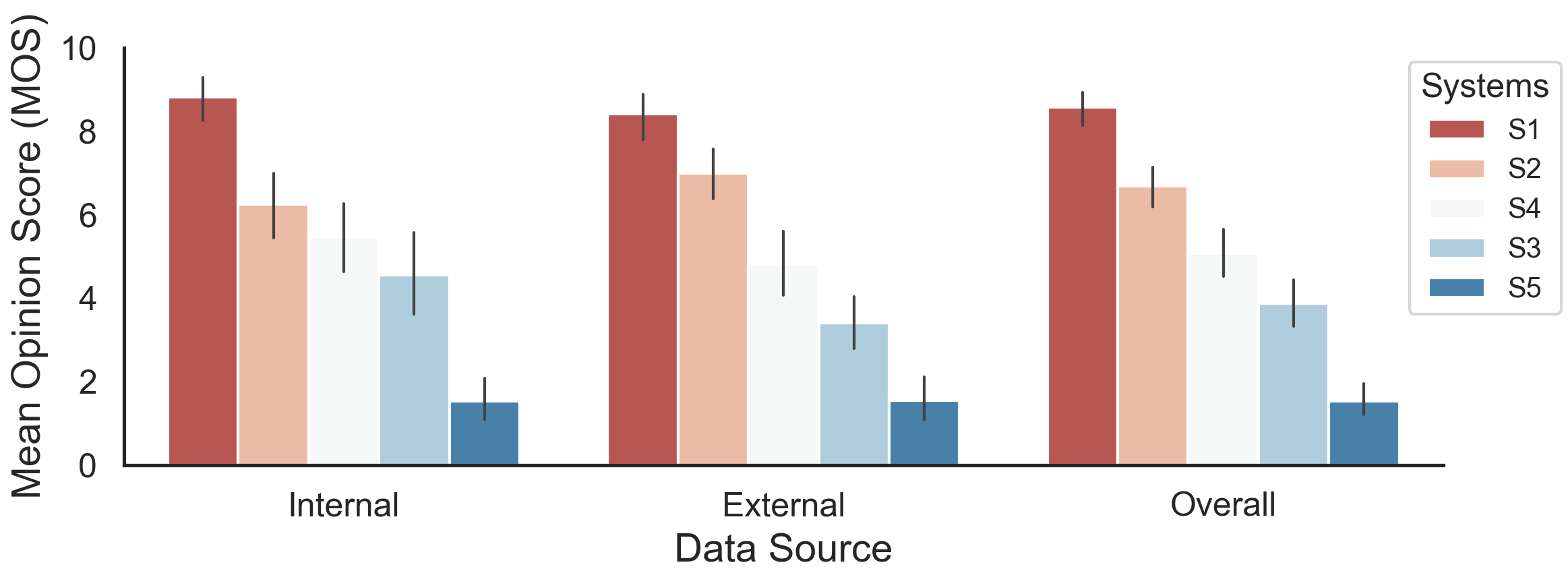}
 \vspace{-1.5em}
  \caption{MOS for systems listed in Table~\ref{tab:listening_test} from Test A. Scores with respect to the internal and external compositions are presented.}
  \label{fig:listening_tests}
  \vspace{-1em}
\end{figure}

We further calculated the MOS for both the external and internal compositions, as shown in Fig.~\ref{fig:listening_tests}. The scores for the M2M outputs synthesised by Pianoteq (\textbf{S2}) demonstrate the M2M model's ability to generalise effectively across compositions with similar styles. However, the proposed system (\textbf{S3}) received noticeably lower scores in both the internal and external groups, indicating the suboptimal performance of the fine-tuned M2A model. Since none of the compositions selected for this test were used during the fine-tuning of the M2A model, these results suggest the model's limited generalisation at this stage.

In the second test (\textbf{Test B}), as shown in Table \ref{tab:listening_test}, participants evaluated the \textit{quality} of the performances, defined by audio clarity and overall sound quality, independent of the musical content. The human performance recordings (\textbf{S0}) were provided as the reference as well. 
Regarding audio quality, the fine-tuned M2A model (\textbf{S7}) received lower ratings compared to the original M2A model (\textbf{S6}). This decrease in quality might come from the acoustic variability in the recordings from the ATEPP dataset, as discussed in Sec.~\ref{sec:fine-tune}. 
Overall, the proposed system demonstrates an improvement in achieving expressiveness and high audio quality, compared to the baseline model, emphasising the need for task-specific considerations.

\section{Discussion}
The results demonstrate that our system effectively recreates human-like expressiveness while preserving acoustic ambience. However, we acknowledge several limitations of the system. First, the inability of the M2M model to generate pedalling adversely affects the M2A model's performance, as the M2A model relies on pedalling to extend note duration when preparing the training data. Nonetheless, the M2A model sometimes compensates by producing meaningful sustain in the notes. Second, the M2M model tends to predict conservative feature values that cluster around the average, lacking distinct stylistic variations across performers. Additionally, the joint training of the M2M and M2A models was not pursued due to the substantial computational resources required by the system design.

\section{Conclusion}
We present an integrated system designed to render expressive piano audio from mechanical MIDI scores. This system combines our proposed M2M model with a fine-tuned MIDI synthesiser, the M2A model. Our evaluation demonstrates that the system effectively reconstructs human-like expressiveness while preserving the acoustic nuances of various recording environments. Although the fine-tuned synthesiser exhibits some limitations in audio quality, the proposed system still achieves improvements over baseline models in producing musical expressiveness and high audio fidelity. Future work will aim to enable pedalling prediction in the M2M model, improve the M2A model by incorporating Chromagram loss to the objective function, and extend its application across various acoustic environments and musical styles.
\newpage
\printbibliography
\end{document}